\documentclass[reprint,amsmath,smsymb,aps,superscriptaddress,aip,apl,nofootinbib]{revtex4-2}
\usepackage{geometry}
\renewcommand{\thesubsection}{\alph{subsection}}
\usepackage{titlesec}
\usepackage{bm}
\usepackage{gensymb}
\usepackage{graphicx}
\usepackage{tensor}
\usepackage{verbatim}
\usepackage{courier}
\usepackage{mathtools}
\usepackage[inline]{enumitem}
\graphicspath{ {./} }
\usepackage{color}
\usepackage{amsmath}
\usepackage{xfrac}
\usepackage[table]{xcolor}
\usepackage{appendix}

\begin{document}
\title{Accelerating Multi-Model Bayesian Inference, Model Selection and Systematic Studies for Gravitational Wave Astronomy}
\author{Charlie Hoy}
\affiliation{Gravity Exploration Institute, School of Physics and Astronomy, Cardiff University, Cardiff, CF24 3AA, United Kingdom}
\date{\today}

\begin{abstract}
Gravitational wave models are used to infer the properties of black holes in merging binaries from the observed gravitational wave signals through Bayesian inference. Although we have access to a large collection of signal models that are sufficiently accurate to infer the properties of black holes, for some signals, small discrepancies in the models lead to systematic differences in the inferred properties. In order to provide a single estimate for the properties of the black holes, it is preferable to marginalize over the model uncertainty. Bayesian model averaging is a commonly used technique to marginalize over multiple models, however, it is computationally expensive. An elegant solution is to simultaneously infer the model and model properties in a joint Bayesian analysis. In this work we demonstrate that a joint Bayesian analysis can not only accelerate but also account for model-dependent systematic differences in the inferred black hole properties. We verify this technique by analysing 100 randomly chosen simulated signals and also the real gravitational wave signal GW200129\_065458. We find that not only do we infer statistically identical properties as those obtained using Bayesian model averaging, but we can sample over a set of three models on average $2.5\times$ faster. In other words, a joint Bayesian analysis that marginalizes over three models takes on average only $20\%$ more time than a single model analysis. We then demonstrate that this technique can be used to accurately and efficiently quantify the support for one model over another, thereby assisting in Bayesian model selection.
\end{abstract}

\maketitle

\section{Introduction}

Since the first detection of gravitational waves (GWs) in 2015~\cite{LIGOScientific:2016aoc},
the Advanced LIGO~\cite{LIGOScientific:2014pky} and
Advanced Virgo~\cite{acernese2014advanced} GW observatories have detected $\sim 90$ GWs originating from compact binary coalescences
(CBCs)~\cite{LIGOScientific:2018mvr, LIGOScientific:2020ibl, LIGOScientific:2021usb, nitz20202, venumadhav2020new, zackay2019highly, zackay2019detecting, Nitz:2021uxj}. The properties
of these binaries are typically inferred through Bayesian
inference~\cite[see e.g.][]{Pankow:2015cra, Ashton:2018jfp, Romero-Shaw:2020owr, Smith:2019ucc, Biwer:2018osg, Lange:2018pyp, Veitch:2014wba, Williams:2021qyt, Gabbard:2019rde, Green:2020dnx, Zackay:2018qdy, Leslie:2021ssu}, where Bayes' theorem is exploited in order to
calculate the
\emph{posterior probability distribution}: the probability that the binary has a specific set of properties given the observed data. In most cases, Bayesian inference is performed by stochastically sampling over the binaries' properties and returning a set of independent samples drawn from the posterior distribution; although see Refs.~\cite{Pankow:2015cra, Lange:2018pyp, Williams:2021qyt, Gabbard:2019rde, Green:2020dnx, Zackay:2018qdy, Leslie:2021ssu, Singer:2015ema, Cornish:2021wxy} for other methods. Two common stochastic sampling approaches are Markov-Chain Monte-Carlo (MCMC)~\cite{metropolis1949monte} and Nested Sampling~\cite{Skilling:2006gxv}.

In order to perform Bayesian inference on GW data, a parameterised model for the GWs emitted from a CBC~\cite{Bohe:2016gbl, Cotesta:2018fcv, Cotesta:2020qhw, Ossokine:2020kjp,Babak:2016tgq,Pan:2013rra, Husa:2015iqa, Khan:2015jqa, London:2017bcn, Hannam:2013oca, Khan:2018fmp, Khan:2019kot, Varma:2019csw, Varma:2018mmi, Pratten:2020fqn, Garcia-Quiros:2020qpx, Pratten:2020ceb, Estelles:2020osj, Estelles:2020twz, Estelles:2021gvs} is
required in order to evaluate the likelihood~\cite{Veitch:2014wba}. It has previously been shown that analyses that use different models can result in noticeably different posterior
distributions for individual observations~\cite[see e.g.][]{Kalaghatgi:2019log,Shaik:2019dym,LIGOScientific:2020ufj,LIGOScientific:2020stg,LIGOScientific:2020zkf,LIGOScientific:2021djp,Hannam:2021pit} and the population as a whole~\cite{Purrer:2019jcp,Moore:2021eok}, which is understood to be caused
by underlying systematic differences in the GW
models. As illustrated most recently in the third GW transient catalog (GWTC-3)~\cite{LIGOScientific:2021djp}, systematic differences in the GW models remains a limiting factor when inferring the properties of black hole binaries (BBHs) through Bayesian inference. Although there are ongoing efforts to enhance the accuracy of future GW models~\cite[see e.g.][]{Hamilton:2021pkf}, systematic differences will likely remain a limiting factor for the foreseeable future.

Bayesian model selection (BMS) provides a method for choosing between posterior distributions obtained with different models. Here, the posterior distribution obtained with the single model that best describes the data is objectively selected, i.e. the model with the largest Bayesian evidence. On the other hand, Bayesian model averaging (BMA) and Bayesian model combination combine posterior distributions obtained with different models to provide a single \emph{model marginalized} result. BMA calculates a weighted average of the posterior distributions obtained with the different models (see Refs.~\cite{LIGOScientific:2018mvr,LIGOScientific:2020ibl,LIGOScientific:2021djp, berry1500597, Ashton:2019leq} and Ref.~\cite{fragoso2018bayesian} for a review) and therefore inherently assumes that only one model is the \emph{true} data generating model but there is an associated uncertainty as to which model it is. Bayesian model combination is similar to BMA but instead assumes that the behaviour of the \emph{true} data generating model can be replicated more closely by a combination of simpler models~\cite{clarke2003comparing}. The posterior distributions obtained with the different models are therefore combined in a variety of ways and the combination is weighted according to the probability that the specific ensemble is correct. Other techniques, including averaging the likelihood for each model at each proposed point during the sampling~\cite{Jan:2020bdz}, have been developed but in general, combining/deciphering between posterior distributions obtained with different models remains a computationally expensive exercise\footnote{We note that although RIFT~\cite{Lange:2018pyp} has implemented a technique to efficiently calculate likelihoods for multiple models since it reduces the computational cost of generating waveforms~\cite{Lange:2018pyp,Wysocki:2019grj}, it remains expensive for
stochastic Bayesian inference methods such as those presented in Refs.~\cite{Ashton:2018jfp, Romero-Shaw:2020owr, Smith:2019ucc, Veitch:2014wba, Biwer:2018osg}.}.

A computationally cheaper solution for marginalizing over the model involves simultaneously inferring the model and model properties in a joint Bayesian analysis (JBA)~\cite{Green:1995mxx}. This has the benefit of ensuring that all models analyse exactly the same data with identical settings, and avoids the need to calculate weights, which are often difficult to calculate robustly. Employing a JBA to marginalize over the model is not a new principle; it has been previously explored both inside~\cite[see e.g.][]{Cornish:2007if,Cornish:2014kda,Littenberg:2014oda,Ashton:2021yum} and outside~\cite[see e.g.][]{andrieu1999joint} of GW research. Specifically, Ref.~\cite{Ashton:2021yum} used a JBA to study model systematics with an emphasis on binary neutron star mergers and the binary neutron star equation of state by using MCMC methods.

In this paper we demonstrate the validity and robustness of using a JBA to marginalize over the model for BBH mergers. We show that a) waveform systematics for BBH mergers can be addressed with a JBA, b) there is a significant reduction in computational cost when using a JBA compared to applying BMA, c) a JBA can accurately and efficiently quantify the support for one model over another, including cases where different models have a different number of parameters to describe additional physics and d) a JBA can be implemented within the Nested Sampling framework. We verified our results by analysing randomly chosen simulated signals and also the real GW signal GW200129\_065458. We show that not only can we obtain statistically identical results compared to applying BMA, but also that the results can be obtained $\sim 2.5\times$ faster when sampling over 3 models. This means that a JBA that marginalizes over 3 models takes on average only $20\%$ more time than a single model analysis. Similarly, we show that a single JBA can produce two distinct Bayes' factors on average $2.6\times$ faster than traditional techniques.

This paper is organized as follows: in Section~\ref{sec:methods} we describe how a JBA can efficiently marginalize over the uncertainty in a set of models. In Section~\ref{sec:validation} we demonstrate that a JBA produces statistically identical results as those obtained when applying BMA. In Section~\ref{sec:200129} we show that a JBA can address model systematics in real GW strain data and highlight that there can be subtleties in its interpretation. In Section~\ref{sec:bayes_factors} we highlight that a JBA has several other applications, including the potential to significantly reduce the computation of one or more Bayes' factors. Finally we conclude with discussions in Section~\ref{sec:discussion}.

\section{Method} \label{sec:methods}

The properties of a system that produced an observed GW, characterised by the multi-dimensional vector $\boldsymbol{\lambda} = \{\lambda_{1}, \lambda_{2}, ..., \lambda_{j}\}$, can be inferred through Bayesian inference. These properties are then represented by the model-dependent posterior distribution $p(\boldsymbol{\lambda} | d, m)$, which is conditional on the observed GW data $d$ and a parameterised model $m$ for the GWs emitted from the system. This model-dependent posterior distribution is calculated using Bayes' theorem,

\begin{equation} \label{eq:bayes_theorem}
    p(\boldsymbol{\lambda} | d, m) = \frac{p(\boldsymbol{\lambda} | m)\, p(d | \boldsymbol{\lambda}, m)}{\mathcal{Z}},
\end{equation}
where $p(\boldsymbol{\lambda} | m)$ is the probability of the system having properties $\boldsymbol{\lambda}$ given the GW model, otherwise known as the prior, $p(d | \boldsymbol{\lambda}, m)$ is the probability of observing the data given the system's properties and GW model, otherwise known as the likelihood, and $\mathcal{Z}$ is the probability of observing the data given the GW model $\mathcal{Z} = p(d | m) \equiv \int{p(\boldsymbol{\lambda} | m)\, p(d | \boldsymbol{\lambda}, m)\, d\boldsymbol{\lambda}}$, otherwise known as the evidence. The model-dependent posterior distribution assumes that model $m$ is correct.

When there is an ensemble of models, $\boldsymbol{m} = \{m_{1}, ..., m_{i},..., m_{N}\}$, BMA can be applied to marginalize over model uncertainty,
\begin{equation} \label{eq:multi_waveform_bayes_theorem}
    p(\boldsymbol{\lambda} | d) = \sum_{i=1}^{N}{p(\boldsymbol{\lambda} | d, m_{i})\, p(m_{i} | d)},
\end{equation}
where $p(m_{i} | d)$ is the model probability (the probability of the model $m_{i}$ given the data) and $N$ is the total number of models that we wish to marginalize over. By exploiting Bayes' theorem, it can be shown that the the model probability is a function of the model's Bayesian evidence~\cite{fragoso2018bayesian,Ashton:2019leq},

\begin{equation} \label{eq:evidence_mixing}
    p(m_{i} | d) = \frac{\mathcal{Z}_{i}\, p(m_{i})}{\sum_{j=1}^{N} \mathcal{Z}_{j}\,p(m_{j})}.
\end{equation}
where $p(m_{i})$ is the prior probability for the choice of model. For the case of uniform priors, $p(m_{i}) = 1/N\,\, \forall\, m_{i}$, BMA is simply the average of the model-dependent posterior distributions weighted by the evidence. The significant disadvantages of using BMA to marginalize over the model is that a) $p(\boldsymbol{\lambda} | d, m_{i})$ must be calculated for all models, which is computationally expensive, and b) $\mathcal{Z}_{i}$ must be robustly inferred which is often difficult to guarantee, especially for GW Bayesian inference where tails of the likelihood surface occupy a large volume in high-dimensions.

An alternative solution for marginalizing over the model is to simultaneously infer the model and model properties in a single JBA~\cite{Green:1995mxx}. For this case, the multi-dimensional vector $\boldsymbol{\lambda}$ can be expanded to include the model, $\tilde{\boldsymbol{\lambda}} = \{\lambda_{1}, \lambda_{2}, ..., \lambda_{j}, m\}$. If the additional degree of freedom representing the choice of model is trivial for existing parameter estimation methods to explore, a JBA will at most be $N\times$ faster to compute compared to applying BMA. For the worst case scenario, where the choice of model is difficult for parameter estimation methods to explore, we expect that a JBA will take the same time to compute as applying BMA.

In general, the JBA returns a single posterior distribution on the multi-dimensional parameter space. The probability distribution for a specific dimension can then be obtained by marginalizing over the unwanted dimensions. Consequently, the model probability can be inferred from the JBA by calculating,

\begin{equation}
    p(m | d) = \int{p(\tilde{\boldsymbol{\lambda}}| d)\, d\lambda_{1}\, d\lambda_{2}\,..., d\lambda_{j}}.
\end{equation}

Although there are multiple approaches for calculating the posterior distribution~\cite[see e.g.][]{Pankow:2015cra, Lange:2018pyp, Veitch:2014wba, Williams:2021qyt, Gabbard:2019rde, Green:2020dnx, Zackay:2018qdy, Leslie:2021ssu, Singer:2015ema, Cornish:2021wxy}, the majority of Bayesian inference analyses use stochastic parameter estimation methods that return a set of discrete samples drawn from the posterior. As explained in Section 5 of Ref.~\cite{Ashton:2021yum}, assuming that the total number of collected samples from a JBA is $\mathcal{N}$, $p(m_{i} | d)$ is simply,

\begin{equation} \label{eq:basic_principle}
    p(m_{i} | d) = \frac{\mathcal{N}_{i}}{\mathcal{N}},
\end{equation}
where $\mathcal{N}_{i}$ is the number of independent samples obtained with model $m_{i}$. Assuming that we have infinite time and perfect sampling, such that we have a) sufficient samples to fully explore the model parameter space and b) a reliable estimate for $\mathcal{Z}_{i}$,

\begin{equation} \label{eq:equivalence}
    \frac{\mathcal{N}_{i}}{\mathcal{N}}\, = \frac{\mathcal{Z}_{i}\, p(m_{i})}{\sum_{j=1}^{N}\mathcal{Z}_{j}\,p(m_{j})}.
\end{equation}
We therefore expect that the JBA will return a similar posterior distribution as that obtained when applying BMA. \footnote{Note that by construction, the independent samples obtained in the JBA are mixed according to the model probabilities. The prior for the model is therefore folded into $\mathcal{N}_{i}$.}.

\section{Implementation and Validation} \label{sec:validation}

A typical Bayesian analysis of a quasicircular BBH analyses a 15-dimensional parameter space $\boldsymbol{\lambda}$: 2 dimensions describing the component masses of the binary ($q, \mathcal{M}$), 6 describing two spin vectors of each component ($a_{1}, a_{2}, \theta_{1}, \theta_{2}, \Delta\phi, \phi_{JL})$, 2 for the binary's inclination and phase ($\theta_{JN}, \phi$), 4 for the binary's location on the sky, distance to the source and polarisation ($\mathrm{RA}, \mathrm{DEC}, d_{\mathrm{L}}, \psi$) and finally 1 describing the merger time of the binary ($t_{c}$). To improve convergence when stochastically sampling over the binaries' properties, the distance to the source is often analytically marginalized over~\cite{Singer:2015ema}, see e.g. Ref.~\cite{LIGOScientific:2021djp}, through the use of a numerical look-up-table. This means that $\boldsymbol{\lambda}$ is often a 14-dimensional parameter space. Consequently, a JBA will analyse a 15-dimensional parameter space $\tilde{\boldsymbol{\lambda}}$: all of the aforementioned dimensions plus an additional dimension describing the model ($m$).

We choose to implement our JBA into the {\sc{Dynesty}} Nested Sampling package~\cite{Speagle:2019ivv} and the {\sc{pBilby}} parameter estimation software (a parallelised version of the {\sc{Bilby}} software~\cite{Ashton:2018jfp, Romero-Shaw:2020owr}) since they are both regularly used by the LIGO-Virgo-KAGRA (LVK) collaborations~\cite[see e.g.][]{LIGOScientific:2020stg, LIGOScientific:2020zkf}. During the sampling of our JBA, a 15-dimensional vector of model parameters is proposed at each step, including an integer representing the model $m$. As was done in Ref.~\cite{Ashton:2021yum}, in order to calculate the likelihood, we first apply a mapping which selects the model based on the integer $m$, and then pass the remaining model parameters to the selected model. We therefore simultaneously explore the model and parameter space.

To validate a) our implementation and b) the robustness of using a JBA to marginalize over model uncertainty for BBH mergers, we perform a comprehensive injection and recovery analysis. We compare the posterior distributions inferred from 100 randomly chosen simulated signals to those obtained when applying BMA to the model-dependent posteriors (see Eq.~\ref{eq:multi_waveform_bayes_theorem}). We also compare the computational cost of both methods.

We choose to inject GWs emitted from 100 randomly chosen BBHs into coloured-Gaussian noise from two detectors, Hanford and Livingston, with design sensitivities~\cite{LIGOScientific:2014pky}. In this validation study we use a selection of \emph{precessing models}, which we define as models that include 6 spin degrees of freedom but do not include higher-order multipoles~\cite{Thorne:1980ru}. We do not include any \emph{complete models}, which we define as models that include all 6 spin degrees of freedom and higher-order multipoles, due to the large computational cost required to analyse 100 simulated signals. We therefore simulate the injected GWs with {\sc{IMRPhenomXP}}~\cite{Pratten:2020ceb} and our recovery analysis marginalizes over {\sc{IMRPhenomXP}}, {\sc{IMRPhenomTP}}~\cite{Estelles:2020osj} and {\sc{IMRPhenomPv3}}~\cite{Khan:2018fmp}.

The parameters of the injected GWs were obtained through random draws of the prior used in the recovery analysis. We used a prior that is uniform over spin magnitudes and component masses, and isotropic over spin orientation, sky location and binary orientation~\cite{LIGOScientific:2018mvr, LIGOScientific:2020ibl, LIGOScientific:2021djp}. The signal-to-noise ratios (SNRs) for the injected GWs ranged from 3.0 to 47.0 with 75\% of the injected binaries having SNR $> 8$ (a typical
search threshold). To improve the convergence of the inferred results, we analytically marginalized over the luminosity distance through the use of a numerical look-up-table. The JBA extended the prior to include an equal-weighted Categorical prior for the choice of model.

\begin{figure}[t!]
    \includegraphics[width=0.52\textwidth]{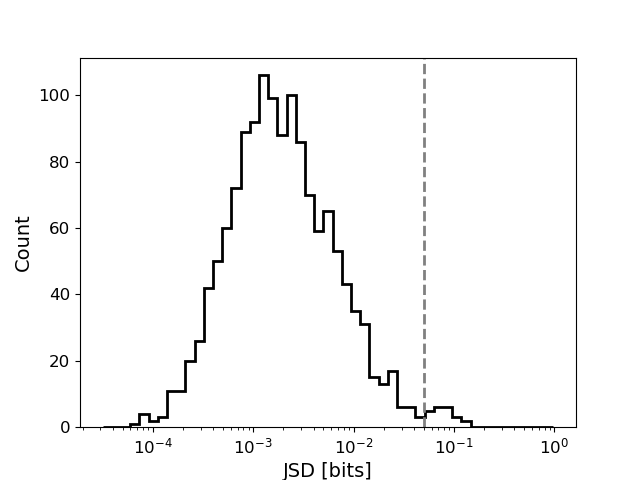}
    \caption{Comparison between the 1400 posterior distributions obtained from the JBA and BMA analyses when analysing 100 randomly chosen binaries. The Jensen-Shannon Divergence (JSD) ranges between $0\,\mathrm{bits}$ and $1\,\mathrm{bits}$ where a JSD=$0\,\mathrm{bits}$ (JSD=$1\,\mathrm{bits}$) implies statistically identical (distinct) distributions. The vertical dashed line shows a JSD$=0.05\,\mathrm{bits}$.}
    \label{fig:comparison}
\end{figure}

For each of the 100 simulated signals, 14 comparable distributions, one for each dimension, can be inferred from both the JBA and BMA analyses. To compare each of these 1400 distributions, we calculate the Jensen-Shannon Divergence (JSD)~\cite{61115}. The JSD ranges between $0\,\mathrm{bits}$ and $1\,\mathrm{bit}$, where a JSD=$0\,\mathrm{bits}$ (JSD=$1\,\mathrm{bit}$) implies statistically identical (distinct) distributions. As shown in Fig.~\ref{fig:comparison}, the inferred JSDs ranged between $7\times10^{-5}\,\mathrm{bits}$ and $0.1\,\mathrm{bits}$ with median and 90\% symmetric credible interval $\mathrm{JSD}=0.002^{+0.010}_{-0.001}\,\mathrm{bits}$. As described in Ref.~\cite{LIGOScientific:2018mvr}, a good rule of thumb is that a JSD $<0.05\,\mathrm{bits}$ implies that the distributions are in good agreement. We find that 98\% of all distributions satisfied the JSD $<0.05\,\mathrm{bits}$ constraint. Consequently, for the majority of cases, there is no statistical difference between the JMA and BMA analyses. Although 2\% of the distributions are statistically different, we find that all distributions are \emph{consistent} between JBA and BMA analyses. This means that the median of the BMA analysis lies within the 90\% confidence interval of the JBA and vice versa. We suspect that these small statistical differences for a subset of distributions are therefore a consequence of sampling uncertainties. We discuss the individual distributions in more detail in Appendix~\ref{sec:pp_test}.

For the 2\% of distributions with JSD $\geq 0.05\,\mathrm{bits}$, we find no correlation between the JSD and the SNR of the injected signal nor the JSD and the probability of a particular model given the data. Indeed, all distributions obtained when analysing the 14 simulated signals with the highest SNRs (ranging between 24.0 and 47.0) and the 28 simulated signals with the lowest SNRs (ranging between 3.0 and 8.0) satisfied the JSD $<0.05\,\mathrm{bits}$ constraint.

The sampling time for a Nested Sampling analysis depends on the stopping criterion; when the stopping criterion is reached, the sampling terminates and independent samples are returned. In order to gauge how computationally expensive the JBA is, we compare the total sampling time taken to analyse 100 simulated signals for both the JBA and BMA analyses when run on identical machines. The total sampling time for the JBA is the total wall clock time taken to generate $p(\tilde{\boldsymbol{\lambda}}| d)$ for each simulated signal. The total sampling time for the BMA analysis on the other hand, is the total wall clock time taken to generate $p(\boldsymbol{\lambda}| d,m_{i})$ for each model in the ensemble for each simulated signal. We found that the total sampling time to perform the JBA for 100 simulated signals was 239.5 hours with each injection taking on average 2.5 hours to complete when parallelised over 200 CPUs. Meanwhile, the total sampling time to apply BMA for 100 simulated signals was 606.5 hours when run on an identical number of CPUs.

We have therefore shown that a JBA is faster than applying BMA with no statistical difference among the obtained posterior distributions. A JBA can therefore successfully marginalize over model uncertainty for BBH mergers. However, one limitation of using a JBA is that it collects fewer independent samples: on average $2\times$ less. In general, if one model has a significantly larger model probability, we expect a comparable number of samples between the JBA and BMA analyses no matter the number of models in the ensemble. However, if all models have comparable probabilities, we expect the JBA to obtain $\sim N\times$ fewer samples than the BMA. Although this may become an issue when the JBA marginalizes over many models with comparable probabilities, this is unlikely to affect typical GW astronomy since a) ordinarily only 2 of the leading models are combined~\cite[see e.g.][]{LIGOScientific:2018mvr,LIGOScientific:2020ibl,LIGOScientific:2021djp} and b) it is unlikely that many models will have comparable model probabilities (see e.g. Sec.~\ref{sec:200129}). Consequently, for typical analyses undertaken by the LVK, a JBA can significantly reduce the computational cost of multi-model Bayesian inference.

\section{Application to Bayesian Model Selection} \label{sec:200129}

The inferred posterior distribution from a Bayesian inference analysis is dependent on the assumed model. Owing to alternative techniques used when first constructing the models~\cite[see e.g.][]{Buonanno:1998gg, Ajith:2007qp, Field:2013cfa}, some models are often more accurate than others in certain regions of the parameter space. For high SNR GW observations, these model discrepancies can show up as systematic differences in the inferred posterior distributions. This was the case for GW200129\_065458~\cite{Hannam:2021pit, LIGOScientific:2021djp, Hu:2022rjq}, hereafter referred to as GW200129, which was an $\mathrm{SNR}\sim 27$ signal generated by the merger of a $\sim 60\,M_{\odot}$ binary~\cite{LIGOScientific:2021djp} and the first signal with strong evidence for spin-induced orbital precession~\cite{Hannam:2021pit}.

One method for dealing with these systematic differences is to independently check the model's accuracy by comparing it to GW signals calculated by numerically solving Einstein's equations for a given binary system~\cite[e.g.][]{Hannam:2021pit}. However, it is often not feasible to apply this method since it is computationally expensive to numerically solve Einstein's equations especially for low-mass binaries. Another method is to use BMS to identify which model has more support in the observed data. BMS objectively selects between different models by identifying the single model that has the largest Bayesian evidence, ${m}^{*} = \mathrm{max}_{m_{i}} p(d|m_{i})$.

As demonstrated in Sec.~\ref{sec:validation}, a JBA returns posterior distributions that are statistically identical to those obtained with BMA in a fraction of the time. This means that the inferred model probabilities from the JBA are consistent with the model's evidence (see Eqs.~\ref{eq:basic_principle} and \ref{eq:equivalence}). We therefore propose that the JBA can be used to accelerate BMS. To demonstrate this, we performed an independent analysis of GW200129 that marginalized over a selection of the latest \emph{complete models}, and inferred the model probability. We chose to include {\sc{NRSur7dq4}} ($m_{\mathrm{NRS}}$)~\cite{Varma:2019csw}, {\sc{IMRPhenomXPHM}} ($m_{\mathrm{XPHM}}$) and {\sc{IMRPhenomTPHM}} ($m_{\mathrm{TPHM}}$)~\cite{Estelles:2021gvs}\footnote{We did not include the {\sc{SEOBNRv4PHM}}~\cite{Ossokine:2020kjp} model owing to the significant computational resources needed to run this model with the {\sc{pBilby}} parameter estimation software. However, it may be possible to include {\sc{SEOBNRv4PHM}} in future studies by using the developments in Ref.~\cite{Gadre:2022sed}.}. For comparison, we also ran 3 additional analyses, one for each model, and inferred the Bayesian evidence. In all analyses, we used the same priors, sampler settings, power spectral densities and calibration envelopes as those described in Ref.~\cite{Hannam:2021pit}. We also used an agnostic prior for the choice of model.

\begin{figure}[t!]
    \includegraphics[width=0.52\textwidth]{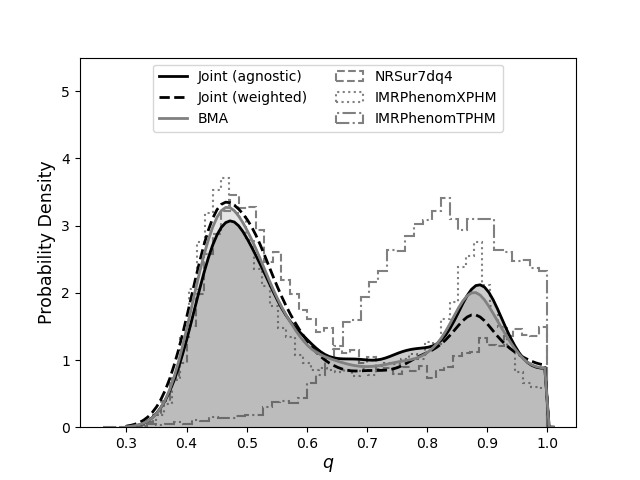}
    \caption{Comparison between the inferred mass ratio $q$, defined as the secondary mass divided by the primary mass and is therefore always $\leq 1$, for GW200129\_065458. The Joint posteriors were obtained by performing a single analysis that marginalized over the {\sc{NRSur7dq4}}, {\sc{IMRPhenomXPHM}} and {\sc{IMRPhenomTPHM}} models. The Joint (agnostic) posterior used a uniform prior for the choice of model and the Joint (weighted) posterior used a prior weighted towards the {\sc{NRSur7dq4}} model. The BMA posterior combined posterior distributions obtained with the {\sc{NRSur7dq4}}, {\sc{IMRPhenomXPHM}}, {\sc{IMRPhenomTPHM}} models according to Eq.~\ref{eq:evidence_mixing}. The JBA was $2.5\times$ faster to compute that the BMA posterior.}
    \label{fig:200129}
\end{figure}

In Fig.~\ref{fig:200129} we show the posterior distribution for the binary's mass ratio, defined as the secondary mass divided by the primary mass $q = m_{2} / m_{1} \leq 1$, inferred from a) the JBA, b) the model-dependent posterior distribution obtained by analysing GW200129 with each model separately, and c) BMA. Although we see significant differences in the inferred model-dependent posterior distributions, the JBA displays excellent agreement with the BMA posterior as expected. This highlights that although $m_{\mathrm{TPHM}}$ suggests that GW200129 is an equal mass ratio system, both the JBA and BMA analyses show that this model is disfavoured in comparison to the $m_{\mathrm{NRS}}$ and $m_{\mathrm{XPHM}}$ models. Although we only plot estimates for the binary's mass ratio, since this is where we saw the biggest discrepancy between the posterior distributions obtained with the individual models when analysing GW200129, we also see excellent agreement between the JBA and BMA analyses for all dimensions. To quantify this agreement, we calculate the JSD between comparable distributions. We found that the JSDs between the JBA and BMA posterior distributions ranged between $0\,\mathrm{bits}$ and $0.009\,\mathrm{bits}$ with median $D_{\mathrm{JS}} = 0.001\,\mathrm{bits}$.

Our JBA inferred model probabilities of 0.30, 0.69 and 0.01 for $m_{\mathrm{NRS}}$, $m_{\mathrm{XPHM}}$ and $m_{\mathrm{TPHM}}$ respectively. According to these model probabilities, $m_{\mathrm{XPHM}}$ best describes the data, $m^{*} = m_{\mathrm{XPHM}}$. This is consistent with the inferred model evidences since the BMA combined individual distributions with weights 0.31, 0.68 and 0.01 for $m_{\mathrm{NRS}}$, $m_{\mathrm{XPHM}}$ and $m_{\mathrm{TPHM}}$ models respectively.

Although the JBA marginalizes over multiple models by construction, we can estimate the model-dependent posterior distribution $p(\boldsymbol{\lambda} | d, m^{*})$ by selecting only the independent samples with model $m^{*}$. Combined with the ability to accurately and efficiently infer the model probabilities, the JBA can be used to accelerate the inference of $p(\boldsymbol{\lambda} | d, m^{*})$.

\begin{figure}[t!]
    \includegraphics[height=0.84\textheight]{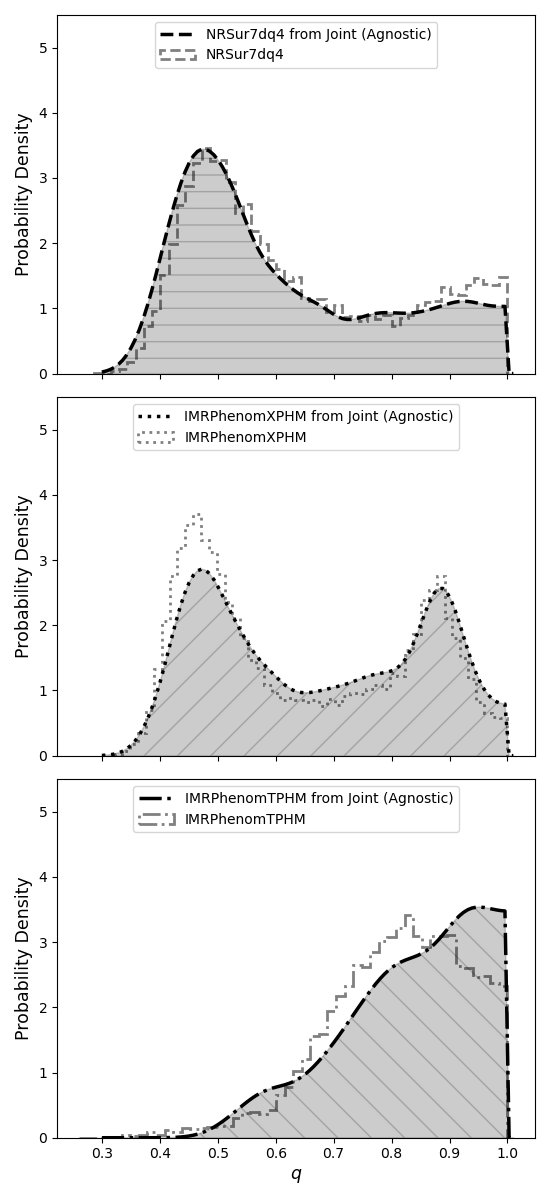}
    \caption{Comparison between the model-dependent posterior for the binaries mass ratio $q$ obtained from analysing GW200129\_065458 with each model separately and an estimate for $p(q | d, m_{i})$ calculated by selecting only the independent samples obtained from the Joint (agnostic) analysis with model $m_{i}$.}
    \label{fig:200129_individual_wvfs}
\end{figure}

Fig.~\ref{fig:200129_individual_wvfs} compares the model-dependent posterior for the binary's mass ratio, $p(q | d, m_{i})$, obtained from analysing GW200129 with each model separately and an estimate for $p(q | d, m_{i})$ calculated by selecting only the independent samples obtained from the JBA with model $m_{i}$. In general we see excellent agreement between the two estimates for $p(q | d, m^{*})$. However, we do see a small difference for $p(q | d, m_{\mathrm{TPHM}})$. While insignificant for a BMS analysis (since $m_{\mathrm{TPHM}}$ has the lowest evidence), this is unsurprising since the JBA posterior contains only $\sim 300$ independent samples with $m_{\mathrm{TPHM}}$ compared to $\sim 44,000$ with $m_{\mathrm{XPHM}}$. Of course, this level of agreement can be improved by allowing the sampler to run for longer and therefore collecting a greater number of samples or by simply combining different independent JBA analyses.

We have therefore shown that a JBA can infer $p(\boldsymbol{\lambda} | d, m_{i})$ and in particular $p(\boldsymbol{\lambda} | d, m^{*})$. As demonstrated previously, a significant advantage of performing the JBA is that the JBA is significantly cheaper to perform. We found that the JBA terminated $2.5\times$ faster than the total time required to apply BMS; the JBA took a total of $\sim 4$ hours to complete on 400 CPUs while applying BMS took a total of $\sim 10$ hours to generate on the same number of CPUs: $\sim 4$ hours for $m_{\mathrm{NRS}}$, $\sim 3.5$ hours for $m_{\mathrm{TPHM}}$ and $\sim 2.5$ for $m_{\mathrm{XPHM}}$ models respectively. As discussed, the JBA obtained $\sim 44,000$ samples for $m^{*}$ while BMS selected the posterior with $\sim 60,000$ samples. In general, the JBA obtained $\sim 20,000$, $\sim 44,000$ and $\sim 300$ samples for $m_{\mathrm{NRS}}$, $m_{\mathrm{XPHM}}$ and $m_{\mathrm{TPHM}}$ respectively while the BMA analysis combined $\sim 30,000$, $\sim 60,000$ and $\sim 1000$ samples for $m_{\mathrm{NRS}}$, $m_{\mathrm{XPHM}}$ and $m_{\mathrm{TPHM}}$ respectively. However, the BMA discarded $\sim 90,000$ model-dependent samples, most obtained with $m_{\mathrm{TPHM}}$.

\subsection{Subtleties in Bayesian Model Selection} \label{sec:subtleties}

Our analysis of GW200129 concludes that $m_{\mathrm{XPHM}}$ has the largest model probability out of the models considered and therefore it best describes the data. At first glance this seems to be inconsistent with the conclusions presented in Ref.~\cite{Hannam:2021pit} since it was found that $m_{\mathrm{NRS}}$ is the most accurate model in the region of parameter space of GW200129. This highlights an important caveat with using the Bayesian evidence for BMS: the model with the larger evidence will not necessarily be the model that most accurately describes the observed GW.

To demonstrate this, consider the following simulated signal: we simulate a signal with parameters that match the most likely GW found from Ref.~\cite{Hannam:2021pit} with $m_{\mathrm{NRS}}$ and inject it into real GW strain data 2 seconds prior to reported merger time of GW200129 in Hanford, Livingston and Virgo. We then analyse the strain data and marginalize over $m_{\mathrm{NRS}}, m_{\mathrm{XPHM}}$ and  $m_{\mathrm{TPHM}}$ using the same settings as described above. Since the ensemble of models used in the recovery analysis includes $m_{\mathrm{NRS}}$, there is no systematic bias as we are including the exact model that was used to simulate the injection; in other words, $m_{\mathrm{NRS}}$ is perfectly accurate for describing the simulated GW.

For this specific simulated signal, our JBA concludes that $m_{\mathrm{XPHM}}$ has the greatest support in the data since the model probabilities are 0.38, 0.58 and 0.04 for $m_{\mathrm{NRS}}$, $m_{\mathrm{XPHM}}$ and $m_{\mathrm{TPHM}}$ models respectively. However, if we compare the maximum likelihood estimates from $m_{\mathrm{NRS}}$ and $m_{\mathrm{XPHM}}$, where the maximum likelihood estimate identifies the parameters $\hat{\boldsymbol{\lambda}}$ that maximizes the likelihood distribution, we find that $m_{\mathrm{NRS}}$ has the larger maximum likelihood. We ascertain that $m_{\mathrm{XPHM}}$ has the greatest support in the data, of those models considered, since it's likelihood distribution is more tightly constrained with a slightly larger prior probability.

This simulated signal highlights that although the model probability can be inferred, this probability does not necessarily correlate with the model accuracy, and therefore care must be taken when interpreting the result. One method for mitigating misinterpretation is to use external knowledge of model accuracy to inform the prior for the choice of model.

\subsection{Priors}

A JBA allows for the user to specify custom priors for the choice of model. This therefore acts as a means of expressing what is already known, or generally agreed upon, regarding the choice of model. One of the problems with using a custom prior is that it must be well motivated.

One option for deriving a custom prior for the choice of model involves generating a weighted Categorical prior, where the weights are correlated with the accuracy of the model, as encoded in the match~\cite[see e.g.][]{Owen:1995tm}, between each model and a series of numerical relativity simulations in the relevant region of parameter space; a model with a larger match has a larger weight. This works on the principle that the match quantifies how similar a waveform is to a fiducial \emph{true} waveform: a match of 1 (0) implies that the two waveforms are identical (orthogonal). For example, the weights could be based on the average value of $\log_{10}\,(1 - \mathrm{match})^{-1}$, since a model with a poor (good) match to numerical relativity simulations will have a smaller (larger) weight; the weights should be normalized such that the sum is unity. However, averaging the match across a given region of parameter space may not be optimal since models tend to perform differently in different regions of the parameter space. Consequently, a parameter space dependent prior may perform better. We leave a detailed investigation to future work.

In order to demonstrate that a BMS analysis can take into consideration external knowledge of model accuracy, we re-analyse the injection detailed in Sec.~\ref{sec:subtleties} with a weighted Categorical prior for the choice of model. To reflect the fact that $m_{\mathrm{NRS}}$ is the most accurate model, we generated a prior that favours $m_{\mathrm{NRS}}$; for the ease of presentation, we chose weights $0.6, 0.3, 0.1$ for $m_{\mathrm{NRS}}$, $m_{\mathrm{XPHM}}$ and $m_{\mathrm{TPHM}}$ respectively. We gave $m_{\mathrm{NRS}}$ twice the weight of $m_{\mathrm{XPHM}}$ since $\log_{10}\,(1 - \mathrm{match})^{-1}$ is roughly twice as large for $m_{\mathrm{NRS}}$ than $m_{\mathrm{XPHM}}$ in the region of parameter space of the simulated signal\footnote{Based on Fig.4 in Ref.~\cite{Hannam:2021pit}, $m_{\mathrm{NRS}}$ and $m_{\mathrm{XPHM}}$ have a match of $\sim 0.999$ and $\sim 0.985$ respectively in the region of parameter space of the simulated signal.}. Under this \emph{informed} prior, the JBA infers model probabilities 0.54, 0.44 and 0.02 for $m_{\mathrm{NRS}}$, $m_{\mathrm{XPHM}}$ and $m_{\mathrm{TPHM}}$ models respectively, which are consistent with simply reweighting the model probabilities obtained under an agnostic prior.

We next use this weighted Categorical prior for a re-analysis of GW200129. In Fig.~\ref{fig:200129} we see that the JBA with a weighted prior more closely resembles the {\sc{NRSur7dq4}} posterior distribution with less support at $q\sim 0.9$ where only the {\sc{IMRPhenomXPHM}} and {\sc{IMRPhenomTPHM}} find any substantial probability. Under the weighted prior, we find that $p(m_{\mathrm{NRS}} | d)$ increases from 0.30 to 0.53. This result is therefore now inline with the conclusions presented in Ref.~\cite{Hannam:2021pit}.

\section{Bayes' factors} \label{sec:bayes_factors}

When multiple models are available, it is natural to quantify how much one model is preferred to another. Within the Bayesian framework, the primary tool for comparing the performance of two competing models is the Bayes' factor. The log Bayes' factor for model $A$ over model $B$ is calculated by simply subtracting the log model evidences: $\log_{10} \mathcal{B}_{AB} = \log_{10}\mathcal{Z}_{A} - \log_{10}\mathcal{Z}_{B}$. A log Bayes' factor $>0$ indicates a preference for model $A$ over model $B$ and in general, if the log Bayes' factor is $>1$ there is strong evidence to suggest that model $A$ is preferred to model $B$~\cite{Kass:1995aaa}.

\begin{figure*}[t!]
    \includegraphics[width=0.99\textwidth]{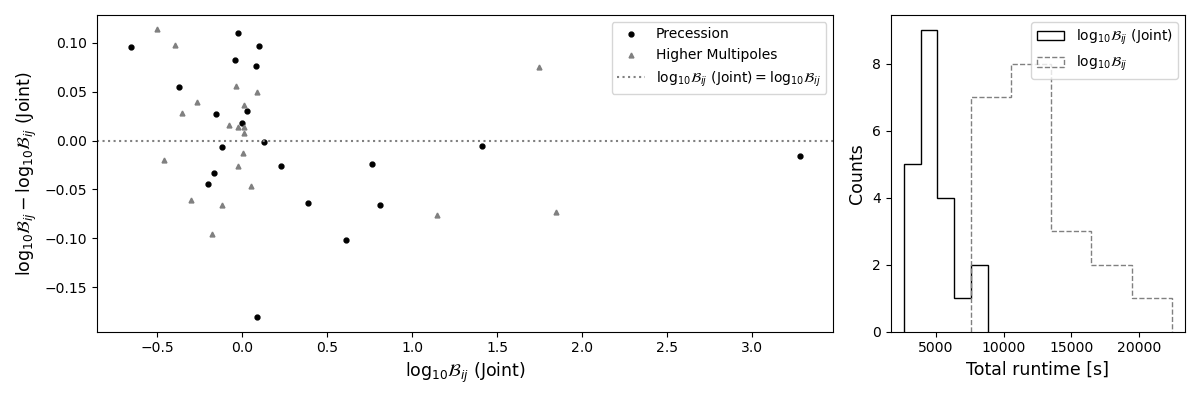}
    \caption{\emph{Left}: Comparison between the $\log_{10}$ Bayes' factors obtained from a single JBA that marginalized over the {\sc{IMRPhenomXAS}}, {\sc{IMRPhenomXP}} and {\sc{IMRPhenomXHM}} models, $\log_{10} \mathcal{B}_{ij}\,\,({\mathrm{Joint}})$, and the $\log_{10}$ Bayes' factors calculated by comparing evidences from 3 separate analyses, an {\sc{IMRPhenomXAS}} analysis, {\sc{IMRPhenomXP}} analysis and an {\sc{IMRPhenomXHM}} analysis, $\log_{10} \mathcal{B}_{ij}$. \emph{Right}: Histogram showing the total time taken to calculate $\log_{10} \mathcal{B}_{ij}\,\,({\mathrm{Joint}})$ and $\log_{10} \mathcal{B}_{ij}$ shown in the \emph{Left} panel.}
    \label{fig:bayes_factors}
\end{figure*}

Since the log Bayes' factor requires estimates for the model's evidence $\mathcal{Z}_{i}$, this requires running multiple parameter estimation analyses, one for each of the models $A$ and $B$, which is computationally expensive. However, as described in Section 5 of Ref.~\cite{Ashton:2021yum}, the log Bayes' factor can be approximated by running a single JBA that marginalizes over models $A$ and $B$. Here, the ratio of inferred model probabilities $p(m_{A} | d) / p(m_{B} | d) = \mathcal{N}_{A} / \mathcal{N}_{B}$ (see Eq.~\ref{eq:basic_principle}) gives the posterior odds for model $A$ over model $B$. For the case of uniform priors for the choice of model, the posterior odds is equivalent to the Bayes' factor,

\begin{equation}
\begin{split}
    \log_{10} \mathcal{B}_{AB} &= \log_{10}\mathcal{Z}_{A} - \log_{10}\mathcal{Z}_{B} \\
    &\approx \log_{10}\mathcal{N}_{A} - \log_{10}\mathcal{N}_{B}.
\end{split}
\end{equation}
If we wish to calculate two log Bayes' factors simultaneously, for example comparing model $A$ to model $B$ and model $A$ to model $C$, we can simply perform a single analysis that marginalizes over all 3 models $A$, $B$ and $C$ while assuming uniform priors for the choice of model, rather than launching 3 separate parameter estimation analyses to calculate $\mathcal{Z}_{A}, \mathcal{Z}_{B}, \mathcal{Z}_{C}$. As demonstrated in Secs.~\ref{sec:validation} and ~\ref{sec:200129}, this has the potential to greatly reduce the computational cost of calculating the Bayes' factor.

In order to show that the JBA can greatly reduce the computational cost of calculating the log Bayes' factor while still maintaining accuracy, we calculated two log Bayes' factors from a single JBA and compared the results to the log Bayes' factors estimated from 3 separate analyses. Owing to the computational cost of generating 2 log Bayes' factors, we only considered the first 20 randomly chosen binaries described in Sec.~\ref{sec:validation}. Unlike in Sec.~\ref{sec:validation}, we simulated the injected GWs with a complete model  ({\sc{IMRPhenomXPHM}}~\cite{Pratten:2020ceb}). Our recovery analysis marginalized over an \emph{aligned-spin} ({\sc{IMRPhenomXAS}}~\cite{Pratten:2020fqn}), precessing ({\sc{IMRPhenomXP}}) and \emph{higher-order multipole} ({\sc{IMRPhenomXHM}}~\cite{Garcia-Quiros:2020qpx}) models.

Since a precessing model includes all 6 spin degrees of freedom, it analyses a 15-dimensional parameter space, see Sec.~\ref{sec:validation} for details. In comparison, an aligned-spin model restricts spins to be aligned with the orbital angular momentum, meaning that only 2 spin degrees of freedom are probed. Consequently an aligned-spin model analyses an 11-dimensional parameter space. The log Bayes' factor in favour of the precessing model over the aligned-spin model therefore quantifies the support for spins misaligned with the orbital angular momentum. This log Bayes' factor can therefore be interpreted as a measure for the evidence of spin-induced orbital precession in the observed GW~\cite{Apostolatos:1994mx}. Similarly, a higher-order multipole model includes higher order multipoles but restricts spins to be aligned with the orbital angular momentum. By calculating the log Bayes' factor in favour of the higher-order multipole model over the aligned-spin model, we quantify the evidence for higher order multipoles in the observed GW\footnote{There are numerous other measures for quantifying the evidence for spin-induced orbital precession and higher-order multipoles in an observed GW, see e.g. Refs.~\cite{Fairhurst:2019srr, Fairhurst:2019vut, Green:2020ptm, Hoy:2021dqg, LIGOScientific:2018mvr,LIGOScientific:2020ibl, LIGOScientific:2020stg,Roy:2019phx, Klimenko:2015ypf, Mills:2020thr}.}.

Our JBA must therefore sample a 15-dimensional parameter space for the precessing model and an 11-dimensional parameter space for both the higher-order multipole and aligned-spin models. Since the aligned-spin parameter space is simply a subspace of the full precessing parameter space, we accomplish this transition by always sampling the 15-dimension parameter space and simply projecting $\tilde{\boldsymbol{\lambda}}$ onto the aligned-spin parameter space when an aligned-spin model is chosen during the sampling. 

The left panel of Fig.~\ref{fig:bayes_factors} compares the log Bayes' factors in favour of precession and higher order multipoles calculated from a single JBA to the log Bayes' factors estimated from 3 separate analyses. In general we see that the difference in log Bayes' factors are $\lesssim 0.1$ with the largest differences occurring for systems with $\log_{10}\mathcal{B}_{ij}\sim 0$.  This level of disagreement is consistent with the expected error in the estimated log Bayes' factor from the {\sc{pBilby}} parameter estimation software and the {\sc{IMRPhenomX}} waveform family, see e.g. Table IV in Ref.~\cite{Colleoni:2020tgc}. We also note that in previous LVK analyses, when $\log_{10}\mathcal{B}_{ij}\sim 0$, the Bayes' factor is not quoted and therefore the difference in $\log_{10}\mathcal{B}_{ij}$ between the two methods is insignificant.

Importantly, we also see that the Bayes' factors are consistent not only for cases where there is very little evidence for precession and/or higher order multipoles ($\sim 0$) but also cases where there is very strong evidence for precession and/or higher order multipoles ($\sim 3.0$). This shows that a JBA can be used to estimate the log Bayes' factor for all signals reported by the LVK to date~\cite{LIGOScientific:2018mvr, LIGOScientific:2020ibl, LIGOScientific:2021djp}. This technique is likely also accurate for signals displaying greater evidence for precession and/or higher order multipoles than $\sim 3.0$ but this region was not explored in our analysis since our binaries were randomly chosen. 

The right panel of Fig.~\ref{fig:bayes_factors} shows the total runtime for computing 40 log Bayes' factors (2 log Bayes' factors for each injection) from a single JBA to the log Bayes' factors estimated from 3 separate analyses. We see that the JBA computed comparable log Bayes' factors on average $2.6\times$ faster than by comparing model evidences. This demonstrates that not only can the JBA maintain accuracy in the computation of the log Bayes' factor, but it can greatly reduce the computational cost.

\section{Discussion} \label{sec:discussion}

In this work we simultaneously inferred the model and model properties in a joint Bayesian analysis. Although JBAs has already been used both inside~\cite[see e.g.][]{Ashton:2021yum} and outside~\cite[see e.g.][]{andrieu1999joint} of GW research, we demonstrated that it can be used to address waveform systematics in BBH mergers while also offering a computationally cheaper solution compared to applying BMA.

We validated this JBA by analysing GWs emitted from 100 randomly chosen binaries where we marginalized over three of the latest GW models. Although our simulated signals had SNRs ranging between 3.0 and 47.0, all of the posterior distributions obtained with the JBA were consistent with those obtained using BMA. In fact, 98\% of the posterior distributions were statistically identical between the two approaches. Remarkably, this method marginalizes over three models on average $2.5\times$ faster than simply applying BMA. As a \emph{real world} example, we also applied the JBA to analyse GW200129\_065458 and investigated how custom priors for the choice of model can be used to reflect our knowledge of model accuracy. We then demonstrated that the model with the largest evidence does not necessarily correlate with the model accuracy and therefore care must be taken when interpreting $p(m|d)$.

This alternative method of marginalizing over a set of models can also be used to significantly reduce the computational cost of computing one or more Bayes' factors. For example, we demonstrated that the Bayes' factor in favour of precession and higher order multipoles can be generated from a single analysis and not only do we obtain comparable Bayes' factors, but we generate them on average $2.6\times$ faster compared to conventional methods.

The method presented in this work is timely since the fourth gravitational-wave observing run, where $\sim 200$ GW signals are likely to be observed, is scheduled to commence early next year. During this observing run we are likely to observe a significant number of events that lie in extreme regions of the parameter space, where it will be desirable to marginalize over the latest waveform models in order to take into consideration model systematics. The method presented in this work provides a simple, robust and computationally efficient way to marginalize over multiple models.

\section{Acknowledgements}
We are grateful to Gregory Ashton, Stephen Fairhurst, Mark Hannam, Soichiro Morisaki, Vivien Raymond and Jonathan Thompson for comments on this manuscript as well as Mark Hannam for continued discussions throughout this project. This work was supported by European Research Council (ERC)
Consolidator Grant 647839. All calculations were performed using the supercomputing facilities at Cardiff University operated by Advanced Research Computing
at Cardiff (ARCCA) on behalf of the Cardiff Supercomputing Facility and the HPC Wales and Supercomputing Wales (SCW) projects. The computational resources at Cardiff University are part-funded by the European Regional Development Fund (ERDF) via the Welsh Government and STFC grant ST/I006285/1. This research made use of data, software and/or web tools obtained from the Gravitational Wave Open Science Center (\href{https://www.gw-openscience.org}{https://www.gw-openscience.org}), a service of LIGO Laboratory, the LIGO Scientific Collaboration and the Virgo Collaboration. LIGO is funded by the U.S. National Science Foundation. Virgo is funded by the French Centre National de Recherche Scientifique (CNRS), the Italian Istituto Nazionale della Fisica Nucleare (INFN) and the Dutch Nikhef, with contributions by Polish and Hungarian institutes. This material is based upon work supported by NSF's LIGO Laboratory which is a major facility fully funded by the National Science Foundation.

Plots were prepared with Matplotlib~\cite{2007CSE.....9...90H}, {\sc{PESummary}}~\cite{Hoy:2020vys} and {\sc{Bilby}}~\cite{Ashton:2018jfp}.
Parameter estimation was performed with the {\sc{pBilby}} parameter estimation software~\cite{Smith:2019ucc}, which made use of the {\sc{dynesty}} nesting sampling package~\cite{Speagle:2019ivv}. {\sc{NumPy}}~\cite{numpy} and {\sc{Scipy}}~\cite{mckinney-proc-scipy-2010} were also used during our analysis.

\appendix

\section{Percentile-Percentile test} \label{sec:pp_test}

\begin{figure}[t!]
    \includegraphics[width=0.47\textwidth]{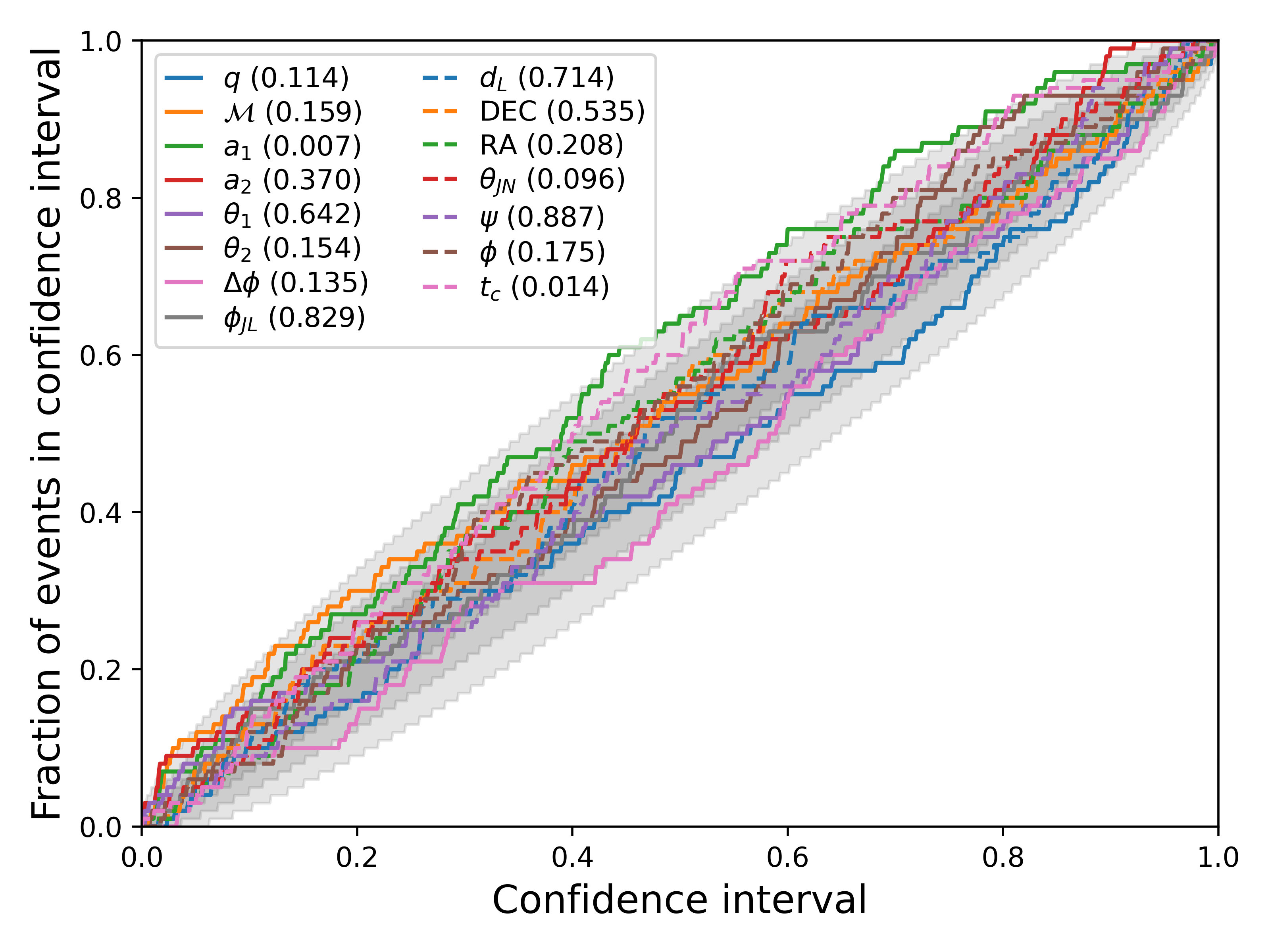}
    \caption{Posterior distributions obtained when analysing 100 randomly chosen binaries modelled with the {\sc{IMRPhenomXP}}~\cite{Pratten:2020ceb} model. The grey regions show the cumulative $1\sigma$, $2\sigma$ and $3\sigma$ confidence intervals expected from a uniform distribution, which is expected if the results are unbiased. Each line shows the cumulative fraction of events within the confidence interval for a given parameter. The posterior distributions were obtained by marginalizing over the {\sc{IMRPhenomXP}}, {\sc{IMRPhenomTP}}~\cite{Estelles:2020osj} and {\sc{IMRPhenomPv3}}~\cite{Khan:2018fmp} models according to Eq.~\ref{eq:basic_principle}. The numbers in the legend show the Kolmogorov-Smirnov p-value~\cite{kolmogorov1933sulla, smirnov1948table} when comparing each line to the expected uniform distribution.}
    \label{fig:pp_test}
\end{figure}

We show the posterior distributions obtained by the JBA described in Section~\ref{sec:validation} in the form of a percentile-percentile (P-P) plot in Fig.~\ref{fig:pp_test}. A P-P plot is useful for identifying biases in the inferred posterior distributions since it plots the fraction of signals with the injected binary lying within a given credible interval against the given credible interval. If the posterior distributions are biased, we would expect to see that the $x\%$ credible interval contains the injected binary more or less than $x\%$ of the time. To quantify the level of bias, we calculate p-values in each dimension, by performing a Kolmogorov-Smirnov (KS) test~\cite{kolmogorov1933sulla, smirnov1948table} against the expected uniform distribution, and a combined p-value, by combining the individual p-values using Fisher's method.

As can be seen in Fig.~\ref{fig:pp_test}, the majority of dimensions show no bias since they are consistent with a uniform distribution at $>5\%$ confidence. Although the primary spin magnitude $a_1$ and coalescence time $t_{c}$ show a small bias (p-values $<0.05$) and the combined p-value shows an overall bias in the inferred distributions: $0.01$, this is unsurprising since the P-P test is only expected to show no bias when the model used to generate the simulated signals exactly matches the model used for recovery~\cite{Ashton:2019leq}. Since we have marginalized over multiple models, each with different inherent assumptions, we have introduced additional uncertainty in the data analysis.

We also generated a P-P plot showing the posterior distributions obtained by the BMA analysis also described in Section~\ref{sec:validation}. As expected, we saw excellent agreement between the JBA and BMA P-P plots. To quantify the level of agreement, we generated individual p-values for each dimension by performing a KS test between comparable lines on the two P-P plots and combined them using Fisher's method. We find a combined p-value of $0.47$, which highlights that there is no statistical difference between P-P plots.

\bibliographystyle{unsrt}
\bibliography{references}

\begin{thebibliography}{10}

\bibitem{LIGOScientific:2016aoc}
B.~P. Abbott et~al.
\newblock {Observation of Gravitational Waves from a Binary Black Hole Merger}.
\newblock {\em Phys. Rev. Lett.}, 116(6):061102, 2016.

\bibitem{LIGOScientific:2014pky}
J.~Aasi et~al.
\newblock {Advanced LIGO}.
\newblock {\em Class. Quant. Grav.}, 32:074001, 2015.

\bibitem{acernese2014advanced}
F~Acernese, M~Agathos, K~Agatsuma, D~Aisa, N~Allemandou, A~Allocca, J~Amarni,
  P~Astone, G~Balestri, G~Ballardin, et~al.
\newblock Advanced virgo: a second-generation interferometric gravitational
  wave detector.
\newblock {\em Classical and Quantum Gravity}, 32(2):024001, 2014.

\bibitem{LIGOScientific:2018mvr}
B.~P. Abbott et~al.
\newblock {GWTC-1: A Gravitational-Wave Transient Catalog of Compact Binary
  Mergers Observed by LIGO and Virgo during the First and Second Observing
  Runs}.
\newblock {\em Phys. Rev. X}, 9(3):031040, 2019.

\bibitem{LIGOScientific:2020ibl}
R.~Abbott et~al.
\newblock {GWTC-2: Compact Binary Coalescences Observed by LIGO and Virgo
  During the First Half of the Third Observing Run}.
\newblock {\em Phys. Rev. X}, 11:021053, 2021.

\bibitem{LIGOScientific:2021usb}
R.~Abbott et~al.
\newblock {GWTC-2.1: Deep Extended Catalog of Compact Binary Coalescences
  Observed by LIGO and Virgo During the First Half of the Third Observing Run}.
\newblock 8 2021.

\bibitem{nitz20202}
Alexander~H Nitz, Thomas Dent, Gareth~S Davies, Sumit Kumar, Collin~D Capano,
  Ian Harry, Simone Mozzon, Laura Nuttall, Andrew Lundgren, and M{\'a}rton
  T{\'a}pai.
\newblock 2-ogc: Open gravitational-wave catalog of binary mergers from
  analysis of public advanced ligo and virgo data.
\newblock {\em The Astrophysical Journal}, 891(2):123, 2020.

\bibitem{venumadhav2020new}
Tejaswi Venumadhav, Barak Zackay, Javier Roulet, Liang Dai, and Matias
  Zaldarriaga.
\newblock New binary black hole mergers in the second observing run of advanced
  ligo and advanced virgo.
\newblock {\em Physical Review D}, 101(8):083030, 2020.

\bibitem{zackay2019highly}
Barak Zackay, Tejaswi Venumadhav, Liang Dai, Javier Roulet, and Matias
  Zaldarriaga.
\newblock Highly spinning and aligned binary black hole merger in the advanced
  ligo first observing run.
\newblock {\em Physical Review D}, 100(2):023007, 2019.

\bibitem{zackay2019detecting}
Barak Zackay, Liang Dai, Tejaswi Venumadhav, Javier Roulet, and Matias
  Zaldarriaga.
\newblock Detecting gravitational waves with disparate detector responses: two
  new binary black hole mergers.
\newblock {\em arXiv preprint arXiv:1910.09528}, 2019.

\bibitem{Nitz:2021uxj}
Alexander~H. Nitz, Collin~D. Capano, Sumit Kumar, Yi-Fan Wang, Shilpa Kastha,
  Marlin Sch\"afer, Rahul Dhurkunde, and Miriam Cabero.
\newblock {3-OGC: Catalog of Gravitational Waves from Compact-binary Mergers}.
\newblock {\em Astrophys. J.}, 922(1):76, 2021.

\bibitem{Pankow:2015cra}
C.~Pankow, P.~Brady, E.~Ochsner, and R.~O'Shaughnessy.
\newblock {Novel scheme for rapid parallel parameter estimation of
  gravitational waves from compact binary coalescences}.
\newblock {\em Phys. Rev. D}, 92(2):023002, 2015.

\bibitem{Ashton:2018jfp}
Gregory Ashton et~al.
\newblock {BILBY: A user-friendly Bayesian inference library for
  gravitational-wave astronomy}.
\newblock {\em Astrophys. J. Suppl.}, 241(2):27, 2019.

\bibitem{Romero-Shaw:2020owr}
I.~M. Romero-Shaw et~al.
\newblock {Bayesian inference for compact binary coalescences with bilby:
  validation and application to the first LIGO\textendash{}Virgo
  gravitational-wave transient catalogue}.
\newblock {\em Mon. Not. Roy. Astron. Soc.}, 499(3):3295--3319, 2020.

\bibitem{Smith:2019ucc}
Rory J.~E. Smith, Gregory Ashton, Avi Vajpeyi, and Colm Talbot.
\newblock {Massively parallel Bayesian inference for transient
  gravitational-wave astronomy}.
\newblock {\em Mon. Not. Roy. Astron. Soc.}, 498(3):4492--4502, 2020.

\bibitem{Biwer:2018osg}
C.~M. Biwer, Collin~D. Capano, Soumi De, Miriam Cabero, Duncan~A. Brown,
  Alexander~H. Nitz, and V.~Raymond.
\newblock {PyCBC Inference: A Python-based parameter estimation toolkit for
  compact binary coalescence signals}.
\newblock {\em Publ. Astron. Soc. Pac.}, 131(996):024503, 2019.

\bibitem{Lange:2018pyp}
Jacob Lange, Richard O'Shaughnessy, and Monica Rizzo.
\newblock {Rapid and accurate parameter inference for coalescing, precessing
  compact binaries}.
\newblock 5 2018.

\bibitem{Veitch:2014wba}
J.~Veitch et~al.
\newblock {Parameter estimation for compact binaries with ground-based
  gravitational-wave observations using the LALInference software library}.
\newblock {\em Phys. Rev. D}, 91(4):042003, 2015.

\bibitem{Williams:2021qyt}
Michael~J. Williams, John Veitch, and Chris Messenger.
\newblock {Nested sampling with normalizing flows for gravitational-wave
  inference}.
\newblock {\em Phys. Rev. D}, 103(10):103006, 2021.

\bibitem{Gabbard:2019rde}
Hunter Gabbard, Chris Messenger, Ik~Siong Heng, Francesco Tonolini, and
  Roderick Murray-Smith.
\newblock {Bayesian parameter estimation using conditional variational
  autoencoders for gravitational-wave astronomy}.
\newblock {\em Nature Phys.}, 18(1):112--117, 2022.

\bibitem{Green:2020dnx}
Stephen~R. Green and Jonathan Gair.
\newblock {Complete parameter inference for GW150914 using deep learning}.
\newblock {\em Mach. Learn. Sci. Tech.}, 2(3):03LT01, 2021.

\bibitem{Zackay:2018qdy}
Barak Zackay, Liang Dai, and Tejaswi Venumadhav.
\newblock {Relative Binning and Fast Likelihood Evaluation for Gravitational
  Wave Parameter Estimation}.
\newblock 6 2018.

\bibitem{Leslie:2021ssu}
Nathaniel Leslie, Liang Dai, and Geraint Pratten.
\newblock {Mode-by-mode relative binning: Fast likelihood estimation for
  gravitational waveforms with spin-orbit precession and multiple harmonics}.
\newblock {\em Phys. Rev. D}, 104(12):123030, 2021.

\bibitem{Singer:2015ema}
Leo~P. Singer and Larry~R. Price.
\newblock {Rapid Bayesian position reconstruction for gravitational-wave
  transients}.
\newblock {\em Phys. Rev. D}, 93(2):024013, 2016.

\bibitem{Cornish:2021wxy}
Neil~J. Cornish.
\newblock {Rapid and Robust Parameter Inference for Binary Mergers}.
\newblock {\em Phys. Rev. D}, 103(10):104057, 2021.

\bibitem{metropolis1949monte}
Nicholas Metropolis and Stanislaw Ulam.
\newblock The monte carlo method.
\newblock {\em Journal of the American statistical association},
  44(247):335--341, 1949.

\bibitem{Skilling:2006gxv}
John Skilling.
\newblock {Nested sampling for general Bayesian computation}.
\newblock {\em Bayesian Analysis}, 1(4):833--859, 2006.

\bibitem{Bohe:2016gbl}
Alejandro Boh\'e et~al.
\newblock {Improved effective-one-body model of spinning, nonprecessing binary
  black holes for the era of gravitational-wave astrophysics with advanced
  detectors}.
\newblock {\em Phys. Rev. D}, 95(4):044028, 2017.

\bibitem{Cotesta:2018fcv}
Roberto Cotesta, Alessandra Buonanno, Alejandro Boh\'e, Andrea Taracchini, Ian
  Hinder, and Serguei Ossokine.
\newblock {Enriching the Symphony of Gravitational Waves from Binary Black
  Holes by Tuning Higher Harmonics}.
\newblock {\em Phys. Rev. D}, 98(8):084028, 2018.

\bibitem{Cotesta:2020qhw}
Roberto Cotesta, Sylvain Marsat, and Michael P\"urrer.
\newblock {Frequency domain reduced order model of aligned-spin
  effective-one-body waveforms with higher-order modes}.
\newblock {\em Phys. Rev. D}, 101(12):124040, 2020.

\bibitem{Ossokine:2020kjp}
Serguei Ossokine et~al.
\newblock {Multipolar Effective-One-Body Waveforms for Precessing Binary Black
  Holes: Construction and Validation}.
\newblock {\em Phys. Rev. D}, 102(4):044055, 2020.

\bibitem{Babak:2016tgq}
Stanislav Babak, Andrea Taracchini, and Alessandra Buonanno.
\newblock {Validating the effective-one-body model of spinning, precessing
  binary black holes against numerical relativity}.
\newblock {\em Phys. Rev. D}, 95(2):024010, 2017.

\bibitem{Pan:2013rra}
Yi~Pan, Alessandra Buonanno, Andrea Taracchini, Lawrence~E. Kidder, Abdul~H.
  Mrou\'e, Harald~P. Pfeiffer, Mark~A. Scheel, and B\'ela Szil\'agyi.
\newblock {Inspiral-merger-ringdown waveforms of spinning, precessing
  black-hole binaries in the effective-one-body formalism}.
\newblock {\em Phys. Rev. D}, 89(8):084006, 2014.

\bibitem{Husa:2015iqa}
Sascha Husa, Sebastian Khan, Mark Hannam, Michael P\"urrer, Frank Ohme, Xisco
  Jim\'enez~Forteza, and Alejandro Boh\'e.
\newblock {Frequency-domain gravitational waves from nonprecessing black-hole
  binaries. I. New numerical waveforms and anatomy of the signal}.
\newblock {\em Phys. Rev. D}, 93(4):044006, 2016.

\bibitem{Khan:2015jqa}
Sebastian Khan, Sascha Husa, Mark Hannam, Frank Ohme, Michael P\"urrer, Xisco
  Jim\'enez~Forteza, and Alejandro Boh\'e.
\newblock {Frequency-domain gravitational waves from nonprecessing black-hole
  binaries. II. A phenomenological model for the advanced detector era}.
\newblock {\em Phys. Rev. D}, 93(4):044007, 2016.

\bibitem{London:2017bcn}
Lionel London, Sebastian Khan, Edward Fauchon-Jones, Cecilio Garc\'\i{}a, Mark
  Hannam, Sascha Husa, Xisco Jim\'enez-Forteza, Chinmay Kalaghatgi, Frank Ohme,
  and Francesco Pannarale.
\newblock {First higher-multipole model of gravitational waves from spinning
  and coalescing black-hole binaries}.
\newblock {\em Phys. Rev. Lett.}, 120(16):161102, 2018.

\bibitem{Hannam:2013oca}
Mark Hannam, Patricia Schmidt, Alejandro Boh\'e, Le\"\i{}la Haegel, Sascha
  Husa, Frank Ohme, Geraint Pratten, and Michael P\"urrer.
\newblock {Simple Model of Complete Precessing Black-Hole-Binary Gravitational
  Waveforms}.
\newblock {\em Phys. Rev. Lett.}, 113(15):151101, 2014.

\bibitem{Khan:2018fmp}
Sebastian Khan, Katerina Chatziioannou, Mark Hannam, and Frank Ohme.
\newblock {Phenomenological model for the gravitational-wave signal from
  precessing binary black holes with two-spin effects}.
\newblock {\em Phys. Rev. D}, 100(2):024059, 2019.

\bibitem{Khan:2019kot}
Sebastian Khan, Frank Ohme, Katerina Chatziioannou, and Mark Hannam.
\newblock {Including higher order multipoles in gravitational-wave models for
  precessing binary black holes}.
\newblock {\em Phys. Rev. D}, 101(2):024056, 2020.

\bibitem{Varma:2019csw}
Vijay Varma, Scott~E. Field, Mark~A. Scheel, Jonathan Blackman, Davide Gerosa,
  Leo~C. Stein, Lawrence~E. Kidder, and Harald~P. Pfeiffer.
\newblock {Surrogate models for precessing binary black hole simulations with
  unequal masses}.
\newblock {\em Phys. Rev. Research.}, 1:033015, 2019.

\bibitem{Varma:2018mmi}
Vijay Varma, Scott~E. Field, Mark~A. Scheel, Jonathan Blackman, Lawrence~E.
  Kidder, and Harald~P. Pfeiffer.
\newblock {Surrogate model of hybridized numerical relativity binary black hole
  waveforms}.
\newblock {\em Phys. Rev. D}, 99(6):064045, 2019.

\bibitem{Pratten:2020fqn}
Geraint Pratten, Sascha Husa, Cecilio Garcia-Quiros, Marta Colleoni, Antoni
  Ramos-Buades, Hector Estelles, and Rafel Jaume.
\newblock {Setting the cornerstone for a family of models for gravitational
  waves from compact binaries: The dominant harmonic for nonprecessing
  quasicircular black holes}.
\newblock {\em Phys. Rev. D}, 102(6):064001, 2020.

\bibitem{Garcia-Quiros:2020qpx}
Cecilio Garc\'\i{}a-Quir\'os, Marta Colleoni, Sascha Husa, H\'ector Estell\'es,
  Geraint Pratten, Antoni Ramos-Buades, Maite Mateu-Lucena, and Rafel Jaume.
\newblock {Multimode frequency-domain model for the gravitational wave signal
  from nonprecessing black-hole binaries}.
\newblock {\em Phys. Rev. D}, 102(6):064002, 2020.

\bibitem{Pratten:2020ceb}
Geraint Pratten et~al.
\newblock {Computationally efficient models for the dominant and subdominant
  harmonic modes of precessing binary black holes}.
\newblock {\em Phys. Rev. D}, 103(10):104056, 2021.

\bibitem{Estelles:2020osj}
H\'ector Estell\'es, Antoni Ramos-Buades, Sascha Husa, Cecilio
  Garc\'\i{}a-Quir\'os, Marta Colleoni, Le\"\i{}la Haegel, and Rafel Jaume.
\newblock {Phenomenological time domain model for dominant quadrupole
  gravitational wave signal of coalescing binary black holes}.
\newblock {\em Phys. Rev. D}, 103(12):124060, 2021.

\bibitem{Estelles:2020twz}
H\'ector Estell\'es, Sascha Husa, Marta Colleoni, David Keitel, Maite
  Mateu-Lucena, Cecilio Garc\'\i{}a-Quir\'os, Antoni Ramos-Buades, and Angela
  Borchers.
\newblock {Time-domain phenomenological model of gravitational-wave subdominant
  harmonics for quasicircular nonprecessing binary black hole coalescences}.
\newblock {\em Phys. Rev. D}, 105(8):084039, 2022.

\bibitem{Estelles:2021gvs}
H\'ector Estell\'es, Marta Colleoni, Cecilio Garc\'\i{}a-Quir\'os, Sascha Husa,
  David Keitel, Maite Mateu-Lucena, Maria de~Lluc Planas, and Antoni
  Ramos-Buades.
\newblock {New twists in compact binary waveform modeling: A fast time-domain
  model for precession}.
\newblock {\em Phys. Rev. D}, 105(8):084040, 2022.

\bibitem{Kalaghatgi:2019log}
Chinmay Kalaghatgi, Mark Hannam, and Vivien Raymond.
\newblock {Parameter estimation with a spinning multimode waveform model}.
\newblock {\em Phys. Rev. D}, 101(10):103004, 2020.

\bibitem{Shaik:2019dym}
Feroz~H. Shaik, Jacob Lange, Scott~E. Field, Richard O'Shaughnessy, Vijay
  Varma, Lawrence~E. Kidder, Harald~P. Pfeiffer, and Daniel Wysocki.
\newblock {Impact of subdominant modes on the interpretation of
  gravitational-wave signals from heavy binary black hole systems}.
\newblock {\em Phys. Rev. D}, 101(12):124054, 2020.

\bibitem{LIGOScientific:2020ufj}
R.~Abbott et~al.
\newblock {Properties and Astrophysical Implications of the 150 M$_\odot$
  Binary Black Hole Merger GW190521}.
\newblock {\em Astrophys. J. Lett.}, 900(1):L13, 2020.

\bibitem{LIGOScientific:2020stg}
R.~Abbott et~al.
\newblock {GW190412: Observation of a Binary-Black-Hole Coalescence with
  Asymmetric Masses}.
\newblock {\em Phys. Rev. D}, 102(4):043015, 2020.

\bibitem{LIGOScientific:2020zkf}
R.~Abbott et~al.
\newblock {GW190814: Gravitational Waves from the Coalescence of a 23 Solar
  Mass Black Hole with a 2.6 Solar Mass Compact Object}.
\newblock {\em Astrophys. J. Lett.}, 896(2):L44, 2020.

\bibitem{LIGOScientific:2021djp}
R.~Abbott et~al.
\newblock {GWTC-3: Compact Binary Coalescences Observed by LIGO and Virgo
  During the Second Part of the Third Observing Run}.
\newblock 11 2021.

\bibitem{Hannam:2021pit}
Mark Hannam, Charlie Hoy, Jonathan~E. Thompson, Stephen Fairhurst, Vivien
  Raymond, and {members of the LIGO, Virgo collaboration}.
\newblock {Measurement of general-relativistic precession in a black-hole
  binary}.
\newblock 12 2021.

\bibitem{Purrer:2019jcp}
Michael P\"urrer and Carl-Johan Haster.
\newblock {Gravitational waveform accuracy requirements for future ground-based
  detectors}.
\newblock {\em Phys. Rev. Res.}, 2(2):023151, 2020.

\bibitem{Moore:2021eok}
Christopher~J. Moore, Eliot Finch, Riccardo Buscicchio, and Davide Gerosa.
\newblock {Testing general relativity with gravitational-wave catalogs: the
  insidious nature of waveform systematics}.
\newblock 3 2021.

\bibitem{Hamilton:2021pkf}
Eleanor Hamilton, Lionel London, Jonathan~E. Thompson, Edward Fauchon-Jones,
  Mark Hannam, Chinmay Kalaghatgi, Sebastian Khan, Francesco Pannarale, and
  Alex Vano-Vinuales.
\newblock {Model of gravitational waves from precessing black-hole binaries
  through merger and ringdown}.
\newblock {\em Phys. Rev. D}, 104(12):124027, 2021.

\bibitem{berry1500597}
Christopher~P.L. {Berry} et~al.
\newblock Quoting parameter-estimation results.
\newblock Technical Report LIGO-T1500597, 2015.

\bibitem{Ashton:2019leq}
Gregory Ashton and Sebastian Khan.
\newblock {Multiwaveform inference of gravitational waves}.
\newblock {\em Phys. Rev. D}, 101(6):064037, 2020.

\bibitem{fragoso2018bayesian}
Tiago~M Fragoso, Wesley Bertoli, and Francisco Louzada.
\newblock Bayesian model averaging: A systematic review and conceptual
  classification.
\newblock {\em International Statistical Review}, 86(1):1--28, 2018.

\bibitem{clarke2003comparing}
Bertrand Clarke.
\newblock Comparing bayes model averaging and stacking when model approximation
  error cannot be ignored.
\newblock {\em Journal of Machine Learning Research}, 4(Oct):683--712, 2003.

\bibitem{Jan:2020bdz}
A.~Z. Jan, A.~B. Yelikar, J.~Lange, and R.~O'Shaughnessy.
\newblock {Assessing and marginalizing over compact binary coalescence waveform
  systematics with RIFT}.
\newblock {\em Phys. Rev. D}, 102(12):124069, 2020.

\bibitem{Wysocki:2019grj}
D.~Wysocki, R.~O'Shaughnessy, Jacob Lange, and Yao-Lung~L. Fang.
\newblock {Accelerating parameter inference with graphics processing units}.
\newblock {\em Phys. Rev. D}, 99(8):084026, 2019.

\bibitem{Green:1995mxx}
Peter~J. Green.
\newblock {Reversible jump Markov chain Monte Carlo computation and Bayesian
  model determination}.
\newblock {\em Biometrika}, 82(4):711--732, 1995.

\bibitem{Cornish:2007if}
Neil~J. Cornish and Tyson~B. Littenberg.
\newblock {Tests of Bayesian Model Selection Techniques for Gravitational Wave
  Astronomy}.
\newblock {\em Phys. Rev. D}, 76:083006, 2007.

\bibitem{Cornish:2014kda}
Neil~J. Cornish and Tyson~B. Littenberg.
\newblock {BayesWave: Bayesian Inference for Gravitational Wave Bursts and
  Instrument Glitches}.
\newblock {\em Class. Quant. Grav.}, 32(13):135012, 2015.

\bibitem{Littenberg:2014oda}
Tyson~B. Littenberg and Neil~J. Cornish.
\newblock {Bayesian inference for spectral estimation of gravitational wave
  detector noise}.
\newblock {\em Phys. Rev. D}, 91(8):084034, 2015.

\bibitem{Ashton:2021yum}
Gregory Ashton and Tim Dietrich.
\newblock {Understanding binary neutron star collisions with hypermodels}.
\newblock {\em Nature Astron.}, 07 2022.

\bibitem{andrieu1999joint}
Christophe Andrieu and Arnaud Doucet.
\newblock Joint bayesian model selection and estimation of noisy sinusoids via
  reversible jump mcmc.
\newblock {\em IEEE Transactions on Signal Processing}, 47(10):2667--2676,
  1999.

\bibitem{Speagle:2019ivv}
Joshua~S. Speagle.
\newblock {dynesty: a dynamic nested sampling package for estimating Bayesian
  posteriors and evidences}.
\newblock {\em Mon. Not. Roy. Astron. Soc.}, 493(3):3132--3158, 2020.

\bibitem{Thorne:1980ru}
K.~S. Thorne.
\newblock {Multipole Expansions of Gravitational Radiation}.
\newblock {\em Rev. Mod. Phys.}, 52:299--339, 1980.

\bibitem{61115}
J.~{Lin}.
\newblock Divergence measures based on the shannon entropy.
\newblock {\em IEEE Transactions on Information Theory}, 37(1):145--151, Jan
  1991.

\bibitem{Buonanno:1998gg}
A.~Buonanno and T.~Damour.
\newblock {Effective one-body approach to general relativistic two-body
  dynamics}.
\newblock {\em Phys. Rev. D}, 59:084006, 1999.

\bibitem{Ajith:2007qp}
Parameswaran Ajith et~al.
\newblock {Phenomenological template family for black-hole coalescence
  waveforms}.
\newblock {\em Class. Quant. Grav.}, 24:S689--S700, 2007.

\bibitem{Field:2013cfa}
Scott~E. Field, Chad~R. Galley, Jan~S. Hesthaven, Jason Kaye, and Manuel
  Tiglio.
\newblock {Fast prediction and evaluation of gravitational waveforms using
  surrogate models}.
\newblock {\em Phys. Rev. X}, 4(3):031006, 2014.

\bibitem{Hu:2022rjq}
Qian Hu and John Veitch.
\newblock {Assessing the model waveform accuracy of gravitational waves}.
\newblock 5 2022.

\bibitem{Gadre:2022sed}
Bhooshan Gadre, Michael P\"urrer, Scott~E. Field, Serguei Ossokine, and Vijay
  Varma.
\newblock {A fully precessing higher-mode surrogate model of effective-one-body
  waveforms}.
\newblock 3 2022.

\bibitem{Owen:1995tm}
Benjamin~J. Owen.
\newblock {Search templates for gravitational waves from inspiraling binaries:
  Choice of template spacing}.
\newblock {\em Phys. Rev. D}, 53:6749--6761, 1996.

\bibitem{Kass:1995aaa}
Robert~E. Kass and Adrian~E. Raftery.
\newblock Bayes factors.
\newblock {\em Journal of the American Statistical Association},
  90(430):773--795, 1995.

\bibitem{Apostolatos:1994mx}
Theocharis~A. Apostolatos, Curt Cutler, Gerald~J. Sussman, and Kip~S. Thorne.
\newblock {Spin induced orbital precession and its modulation of the
  gravitational wave forms from merging binaries}.
\newblock {\em Phys. Rev. D}, 49:6274--6297, 1994.

\bibitem{Fairhurst:2019srr}
Stephen Fairhurst, Rhys Green, Mark Hannam, and Charlie Hoy.
\newblock {When will we observe binary black holes precessing?}
\newblock {\em Phys. Rev. D}, 102(4):041302, 2020.

\bibitem{Fairhurst:2019vut}
Stephen Fairhurst, Rhys Green, Charlie Hoy, Mark Hannam, and Alistair Muir.
\newblock {Two-harmonic approximation for gravitational waveforms from
  precessing binaries}.
\newblock {\em Phys. Rev. D}, 102(2):024055, 2020.

\bibitem{Green:2020ptm}
Rhys Green, Charlie Hoy, Stephen Fairhurst, Mark Hannam, Francesco Pannarale,
  and Cory Thomas.
\newblock {Identifying when Precession can be Measured in Gravitational
  Waveforms}.
\newblock {\em Phys. Rev. D}, 103(12):124023, 2021.

\bibitem{Hoy:2021dqg}
Charlie Hoy, Cameron Mills, and Stephen Fairhurst.
\newblock {Evidence for subdominant multipole moments and precession in merging
  black-hole-binaries from GWTC-2.1}.
\newblock {\em Phys. Rev. D}, 106(2):023019, 2022.

\bibitem{Roy:2019phx}
Soumen Roy, Anand~S. Sengupta, and K.~G. Arun.
\newblock {Unveiling the spectrum of inspiralling binary black holes}.
\newblock {\em Phys. Rev. D}, 103(6):064012, 2021.

\bibitem{Klimenko:2015ypf}
S.~Klimenko et~al.
\newblock {Method for detection and reconstruction of gravitational wave
  transients with networks of advanced detectors}.
\newblock {\em Phys. Rev. D}, 93(4):042004, 2016.

\bibitem{Mills:2020thr}
Cameron Mills and Stephen Fairhurst.
\newblock {Measuring gravitational-wave higher-order multipoles}.
\newblock {\em Phys. Rev. D}, 103(2):024042, 2021.

\bibitem{Colleoni:2020tgc}
Marta Colleoni, Maite Mateu-Lucena, H\'ector Estell\'es, Cecilio
  Garc\'\i{}a-Quir\'os, David Keitel, Geraint Pratten, Antoni Ramos-Buades, and
  Sascha Husa.
\newblock {Towards the routine use of subdominant harmonics in
  gravitational-wave inference: Reanalysis of GW190412 with generation X
  waveform models}.
\newblock {\em Phys. Rev. D}, 103(2):024029, 2021.

\bibitem{2007CSE.....9...90H}
J.~D. {Hunter}.
\newblock {Matplotlib: A 2D Graphics Environment}.
\newblock {\em CSE}, 9:90--95, May 2007.

\bibitem{Hoy:2020vys}
Charlie Hoy and Vivien Raymond.
\newblock {PESummary: the code agnostic Parameter Estimation Summary page
  builder}.
\newblock {\em SoftwareX}, 15:100765, 2021.

\bibitem{numpy}
Oliphant Travis~E.
\newblock {}a guide to numpy, 2006.

\bibitem{mckinney-proc-scipy-2010}
Wes McKinney.
\newblock Data structures for statistical computing in python.
\newblock In St\'efan van~der Walt and Jarrod Millman, editors, {\em
  Proceedings of the 9th Python in Science Conference}, pages 51 -- 56, 2010.

\bibitem{kolmogorov1933sulla}
Andrey Kolmogorov.
\newblock Sulla determinazione empirica di una lgge di distribuzione.
\newblock {\em Inst. Ital. Attuari, Giorn.}, 4:83--91, 1933.

\bibitem{smirnov1948table}
Nickolay Smirnov.
\newblock Table for estimating the goodness of fit of empirical distributions.
\newblock {\em The annals of mathematical statistics}, 19(2):279--281, 1948.

\end{thebibliography}

\end{document}